\documentclass[aps,preprint]{revtex4}%
\usepackage{amsfonts}
\usepackage{amsmath}
\usepackage{amssymb}
\usepackage{graphicx}%
\begin{document}
\title{{\LARGE GRAVITONS, DARK MATTER, AND CLASSICAL GRAVITATION}}
\author{Marcelo Samuel Berman$^{1}$}
\affiliation{$^{1}$Instituto Albert Einstein / Latinamerica\ - Av. Candido Hartmann, 575 -
\ \# 17}
\affiliation{80730-440 - Curitiba - PR - Brazil email: msberman@institutoalberteinstein.org}
\date{(Last Version: 16 July, 2009)}

\begin{abstract}
We find that the quantum of gravity, the \ \textit{graviton}\ \ , has
time-varying mass (the \textit{gomidium}), and radius (the \textit{somium});
both vary with \ \ $R^{-1}$\ \ ; and its frequency is given by Hubble's
parameter. Dark matter can be made of such gravitons. The number of gravitons
varies with \ $R^{2}$\ .\ The Machian radiated power of the Universe leads us
to a "correct" energy density for the gravitons as dark matter.

\end{abstract}
\maketitle

{\LARGE GRAVITONS, DARK MATTER, AND CLASSICAL GRAVITATION }

\begin{center}
{\large MARCELO SAMUEL BERMAN }
\end{center}

\bigskip

\bigskip

\bigskip Berman (2007, 2007a, 2007b, 2008), has suggested that the relative
contributions to the total energy of the Universe, as composed by inertial
mass-energy, gravitational constant, and spin of the Machian Universe, rest
unchanged through its history.

\bigskip

\bigskip According to Heisenberg's uncertainty principle, any two conjugate
quantities, in the sense of Hamilton's canonical ones, carry uncertainties,
\ $\Delta Q$\ \ and \ \ $\Delta P$\ \ , which obey the condition (Leighton, 1959),

\bigskip

$\Delta Q\Delta P\gtrapprox h$\ \ \ \ \ \ \ \ \ \ \ \ \ \ \ \ \ .\ \ \ \ \ \ \ \ \ \ \ \ \ \ \ \ \ \ \ \ \ \ \ \ \ \ \ \ \ \ \ \ \ \ \ \ \ \ \ \ \ \ \ \ \ \ \ \ \ \ \ \ \ \ \ (1)

\bigskip

If we consider maxima \ $\Delta P$\ , we obtain minima \ $\Delta Q$\ \ .

\bigskip

If \ $\Delta P$\ \ stands for the uncertainty in linear momentum, given, say,
by the product of mass and speed, then, its maximum value must be the product
of the largest mass in the Universe by the largest speed,

\bigskip

$\Delta P=M_{U}$ $c$ \ \ \ \ \ \ \ \ \ \ \ \ \ \ . \ \ \ \ \ \ \ \ \ \ \ \ \ \ \ \ \ \ \ \ \ \ \ \ \ \ \ \ \ \ \ \ \ \ \ \ \ \ \ \ \ \ \ \ \ \ \ \ \ \ \ \ \ \ \ \ (2)

\bigskip

We thus, obtain a minimum length value,

\bigskip

$\Delta Q\approx\frac{h}{c\text{ }M_{U}\text{ }}$
\ \ \ \ \ \ \ \ \ \ \ \ \ \ \ \ . \ \ \ \ \ \ \ \ \ \ \ \ \ \ \ \ \ \ \ \ \ \ \ \ \ \ \ \ \ \ \ \ \ \ \ \ \ \ \ \ \ \ \ \ \ \ \ \ \ \ \ \ \ \ \ (3)

\bigskip

\bigskip When we plug the \ mass of the Universe \ $10^{56}$\ grams , we find,
what we call the minimal length in the present Universe ( \ \textit{somium}%
\ \ , after M.M. Som):

\bigskip

$\Delta Q\approx\Delta l=10^{-92}$\ cm\ \ \ \ \ \ \ \ \ \ \ \ . \ \ \ \ \ \ \ \ \ \ \ \ \ \ \ \ \ \ \ \ \ \ \ \ \ \ \ \ \ \ \ \ \ \ \ \ \ \ \ \ \ \ \ \ \ \ (4)

\bigskip

Now, let us think of the largest time (age) in the Universe,

\bigskip

$t_{U}\approx10^{10}$ \ \ years. \ \ \ \ \ \ \ \ \ \ \ \ \ \ \ \ \ \ \ \ \ \ \ \ \ \ \ \ \ \ \ \ \ \ \ \ \ \ \ \ \ \ \ \ \ \ \ \ \ \ \ \ \ \ \ \ \ \ \ \ \ \ \ \ (5)

\bigskip

Its conjugate variable, will point out to a minimum inertial energy and a
minimum inertial mass ( $\Delta m$\ )\ \ \ , \ \ 

$\bigskip$

$\Delta E=c^{2}\Delta m\approx\frac{h}{t_{U}}$\ \ \ \ \ \ \ \ \ \ \ \ \ \ \ . \ \ \ \ \ \ \ \ \ \ \ \ \ \ \ \ \ \ \ \ \ \ \ \ \ \ \ \ \ \ \ \ \ \ \ \ \ \ \ \ \ \ \ \ \ \ \ \ \ \ \ (6)

\bigskip

\bigskip We find, then, the minimal mass in the Universe, which we associate
with the mass of present day quantum, which we term the \ \ \textit{gomidium}%
\ \ (after F.M. Gomide):

\bigskip

$\Delta m\approx10^{-65}$ \ \ grams \ \ \ \ \ \ \ \ \ \ . \ \ \ \ \ \ \ \ \ \ \ \ \ \ \ \ \ \ \ \ \ \ \ \ \ \ \ \ \ \ \ \ \ \ \ \ \ \ \ \ \ \ \ \ \ \ \ \ \ \ \ \ (7)

\bigskip

Gomide (1963) has found the above mass, in other schema.

Analogously, we could repeat the calculation for Planck's mass, and we then
would obtain numerically the same values attained by \ \ \textit{gomidia}%
\ \ and \ \ \textit{somia} \ in Planck's Universe, which coincide with
Planck's length and Planck's\ \ time:

\bigskip

$M_{Pl}c\Delta l_{Pl}=\frac{h}{2\pi}$\ \ \ \ \ \ \ \ \ \ \ \ \ \ \ \ \ ,\ \ \ \ \ \ \ \ \ \ \ \ \ \ \ \ \ \ \ \ \ \ \ \ \ \ \ \ \ \ \ \ \ \ \ \ \ \ \ \ \ \ \ \ \ \ \ \ \ \ \ \ (8)

\bigskip

\bigskip$M_{Pl}c^{2}\Delta t_{Pl}=\frac{h}{2\pi}$ \ \ \ \ \ \ \ \ \ \ \ \ \ \ .\ \ \ \ \ \ \ \ \ \ \ \ \ \ \ \ \ \ \ \ \ \ \ \ \ \ \ \ \ \ \ \ \ \ \ \ \ \ \ \ \ \ \ \ \ \ \ \ \ \ \ \ (9)

\bigskip The two above relations, arise from their respective Planck definitions.

\bigskip

These are the values of Planck's time \ \textit{gomidium}\ \ ( $M_{Pl}$\ ) ,
and \ \ \textit{somium}\ \ ( \ $\Delta l_{Pl}$\ \ ).\ We must yet put a law
for the time variation of \ \textit{gomidia}\ \ \ and \ \textit{somia}\ \ . We
recognize that the relation between the \ \ \textit{gomidium}\ \ values for
the present Universe, and for Planck's one, are related by:

\bigskip

$\frac{\Delta m}{M_{Pl}}\approx\left(  \frac{R}{\Delta l_{Pl}}\right)
^{-1}\approx10^{-60}$\ \ \ \ \ \ \ \ \ \ \ \ \ \ \ \ \ .\ \ \ \ \ \ \ \ \ \ \ \ \ \ \ \ \ \ \ \ \ \ \ \ \ \ \ \ \ \ \ \ \ \ \ \ \ \ \ \ (10)

\bigskip On the other hand, we find the ratio between the \ \textit{somium}%
\ \ \ for the present Universe, with the \ Planck's length,

\bigskip

$\frac{\Delta l}{\Delta l_{Pl}}\approx\frac{\Delta l_{Pl}}{R}\approx10^{-60}$\ \ \ \ \ \ \ \ \ \ \ \ \ \ \ .\ \ \ \ \ \ \ \ \ \ \ \ \ \ \ \ \ \ \ \ \ \ \ \ \ \ \ \ \ \ \ \ \ \ \ \ \ \ \ \ \ \ \ \ \ \ \ \ \ (11)

\bigskip

\bigskip We have found that the \ \textit{gomidium }\ and \ the
\ \ \textit{somium}\ \ , depend on \ $R^{-1}$\ . Consider now that the
Universe has a spin, \ $L$\ \ (Berman, 2007b, 2008, 2008a, 2008b). For
Planck's time, we have only one possibility,

\bigskip

$L_{Pl}=\frac{h}{2\pi}$\ \ \ \ \ \ \ \ \ \ .\ \ \ \ \ \ \ \ \ \ \ \ \ \ \ \ \ \ \ \ \ \ \ \ \ \ \ \ \ \ \ \ \ \ \ \ \ \ \ \ \ \ \ \ \ \ \ \ \ \ \ \ \ \ \ \ \ \ \ \ \ \ \ \ \ \ \ \ \ (12)

\bigskip

An expression for the spin of the Universe, at any time, that would reduce to
the above Planck's value for Planck's time, is given by:

\bigskip

$L=MRc$ \ \ \ \ \ \ \ \ \ \ \ .\ \ \ \ \ \ \ \ \ \ \ \ \ \ \ \ \ \ \ \ \ \ \ \ \ \ \ \ \ \ \ \ \ \ \ \ \ \ \ \ \ \ \ \ \ \ \ \ \ \ \ \ \ \ \ \ \ \ \ \ \ \ \ \ \ \ \ (13)

\bigskip

Berman (2007, 2007a, 2008) has shown that as a result of the Machian
hypothesis above, the radius of the Universe would be linearly proportional to
its mass. This being the case, the last relation implies that,

\bigskip

$L\propto R^{2}$\ \ \ \ \ \ \ \ \ \ \ \ \ \ \ \ .\ \ \ \ \ \ \ \ \ \ \ \ \ \ \ \ \ \ \ \ \ \ \ \ \ \ \ \ \ \ \ \ \ \ \ \ \ \ \ \ \ \ \ \ \ \ \ \ \ \ \ \ \ \ \ \ \ \ \ \ \ \ \ \ \ \ (14)

\bigskip

Consider now that all the gravitons of cosmological origin, are aligned with
the spin of the Universe, due to a universal magnetic field. If the number of
such particles performs a total spin given by \ \ $L$\ \ \ , we would find the
number of such particles,

\bigskip

$n\propto R^{2}$ \ \ \ \ \ \ \ \ \ \ \ \ \ \ .\ \ \ \ \ \ \ \ \ \ \ \ \ \ \ \ \ \ \ \ \ \ \ \ \ \ \ \ \ \ \ \ \ \ \ \ \ \ \ \ \ \ \ \ \ \ \ \ \ \ \ \ \ \ \ \ \ \ \ \ \ \ \ \ \ \ (15)

\bigskip

\bigskip From the above, we find the energy density of the hypothetical gravitons,

\bigskip

$\rho=\frac{n\Delta m}{\frac{4}{3}\pi R^{3}}\propto R^{-2}$ \ \ \ \ \ \ \ \ \ \ \ .\ \ \ \ \ \ \ \ \ \ \ \ \ \ \ \ \ \ \ \ \ \ \ \ \ \ \ \ \ \ \ \ \ \ \ \ \ \ \ \ \ \ \ \ \ \ \ \ \ \ \ \ \ \ (16)

\bigskip

We hint that the gravitons with such mass, will perform the task of dark
matter in the Universe. The numerical values of \ \ $\rho$\ \ \ , \ can be
compared with the inertial density in the Universe,

\bigskip

$\rho_{i}=\frac{M}{\frac{4}{3}\pi R^{3}}\propto R^{-2}$ \ \ \ \ \ \ \ \ \ \ \ .\ \ \ \ \ \ \ \ \ \ \ \ \ \ \ \ \ \ \ \ \ \ \ \ \ \ \ \ \ \ \ \ \ \ \ \ \ \ \ \ \ \ \ \ \ \ \ \ \ \ \ \ (17)

\bigskip

If \ \ $\rho_{i}$\ \ \ amounts to \ $5\%$ \ of the critical energy, and
\ $\rho$\ \ \ equals \ \ $27\%$\ \ , the same \ \ $R^{-2}$\ \ dependence,
makes the relative contributions from dark matter (i.e. the gravitons)
and\ \ visible mass (inertial), become constant over history.

\bigskip

\bigskip If the graviton has a frequency \ $\nu$\ \ , its energy is given by,

\bigskip

$E=\frac{h}{2\pi}\nu=\Delta mc^{2}$\ \ \ \ \ \ \ \ \ \ \ \ \ \ .\ \ \ \ \ \ \ \ \ \ \ \ \ \ \ \ \ \ \ \ \ \ \ \ \ \ \ \ \ \ \ \ \ \ \ \ \ \ \ \ \ \ \ \ \ \ \ \ \ \ \ \ \ (18)

\bigskip

For the present Universe, we would find:

\bigskip

$\nu\approx10^{-18}$ sec$^{-1}$ \ \ \ \ \ \ \ \ \ \ \ \ .\ \ \ \ \ \ \ \ \ \ \ \ \ \ \ \ \ \ \ \ \ \ \ \ \ \ \ \ \ \ \ \ \ \ \ \ \ \ \ \ \ \ \ \ \ \ \ \ \ \ \ \ \ \ \ \ \ (19)

\bigskip

This makes for Hubble's constant.

\bigskip

Consider now the Machian picture of the cosmological gravitons energy density.
From the theory of gravitational radiation, we know that, for a spinning
"sphere", oblated, the power of radiation is given by Einstein's quadrupole formula,

\bigskip

$P=\frac{dE}{dt}=-\frac{G}{5c^{5}}\sum\dddot{Q}_{\alpha\beta}^{2}$
\ \ \ \ \ \ \ \ \ \ \ \ \ \ \ \ \ \ \ \ \ \ \ , \ \ \ \ \ \ \ \ \ \ \ \ \ \ \ \ \ \ \ \ \ \ \ \ \ \ \ \ \ \ \ \ \ \ \ \ \ \ \ (20)

\bigskip

where \ $\dddot{Q}_{\alpha\beta}^{2}$\ \ stands for the quadrupole moment of
the mass distribution.\bigskip\ For a periodic rotation, with Machian angular
speed, \ 

$\bigskip$

$\omega=\frac{c}{R}$\ \ \ \ \ \ \ \ \ \ \ \ \ \ \ \ \ \ \ \ \ \ \ \ , \ \ \ \ \ \ \ \ \ \ \ \ \ \ \ \ \ \ \ \ \ \ \ \ \ \ \ \ \ \ \ \ \ \ \ \ \ \ \ \ \ \ \ \ \ \ \ \ \ \ \ \ \ \ \ \ \ \ \ \ \ \ \ (21)

\bigskip

relation (20) will be given by,

\bigskip

$P\approx\frac{c^{5}}{G}\left(  \frac{GM}{c^{2}R}\right)  ^{2}\approx
\frac{c^{5}}{G}=10^{52}$ watts\ \ \ \ \ \ \ \ \ \ \ \ \ \ \ \ \ , \ \ \ \ \ \ \ \ \ \ \ \ \ \ \ \ \ \ \ \ \ \ \ \ \ \ \ \ (22)

\bigskip

because of the Brans-Dicke relation \ \ $\frac{GM}{c^{2}R}\approx1$\ \ \ \ .\ \ \ \ \ 

\bigskip

\bigskip Again, for a Machian Universe, we may take the radius definition,

\bigskip

$R\approx cT$ \ \ \ \ \ \ \ \ \ \ \ \ \ \ \ \ , \ \ \ \ \ \ \ \ \ \ \ \ \ \ \ \ \ \ \ \ \ \ \ \ \ \ \ \ \ \ \ \ \ \ \ \ \ \ \ \ \ \ \ \ \ \ \ \ \ \ \ \ \ \ \ \ \ \ \ \ \ \ \ \ \ \ \ \ \ \ (23)

\bigskip

where \ $T$\ \ is Hubble's time; we find then, the mass density,

\bigskip

$\rho=\frac{c^{4}}{G}R^{-2}$\ \ \ \ \ \ \ \ \ \ \ \ \ \ \ \ \ \ . \ \ \ \ \ \ \ \ \ \ \ \ \ \ \ \ \ \ \ \ \ \ \ \ \ \ \ \ \ \ \ \ \ \ \ \ \ \ \ \ \ \ \ \ \ \ \ \ \ \ \ \ \ \ \ \ \ \ \ \ \ \ \ \ (24)

\bigskip

If we plug \ \ $R\cong10^{28}$\ cm\ , we shall find that the mass energy
density of the cosmological gravitons is given by the correct order of
magnitude of dark matter density.

Summarizing, Heisenberg's uncertainty principle, yields the quanta of length
and mass, respectively, the \ \ \textit{somium }\ \ and \ \ \textit{gomidium}%
\ \ , which we associate with the graviton, and dark matter. For Planck's
Universe, the minimal length, mass and spin, are given by Planck's values. We
are in face of time-varying gravitons. The mass of such gravitons, may not be
rest one, but the equivalent of mass-energy given by the product of Planck's
constant and a frequency characteristic of the graviton, given by Hubble's
constant for the present Universe.\ \ \ 

\bigskip

\bigskip{\Large Acknowledgments}

\bigskip

The author expresses his recognition to his intellectual mentors, friends and
colleagues, M.M. Som (deceased) and F.M. Gomide. He thanks the many other
colleagues that collaborate with him. The typing was made by Marcelo F.
Guimar\~{a}es, who I consider a friend and to whom my thanks are due for this
and many other collaborations. And, last, but not least, I mention the
encouragement by Albert, Paula and Geni.

\bigskip\bigskip

{\Large References}

\bigskip

\bigskip Berman,M.S. (2007) - \textit{Introduction to General Relativity, and
the Cosmological Constant Problem}, Nova Science, New York..

Berman,M.S. (2007a) - \textit{Introduction to General Relativistic and
Scalar-Tensor Cosmologies}, Nova Science, New York.

Berman,M.S. (2007b) - \textit{The Pioneer anomaly and a Machian Universe.
}Astrophysics and Space Science, \textbf{312}, 275. See a previous version in
Los Alamos Archives, http://arxiv.org/abs/physics/0606117

Berman,M.S. (2008) - \textit{A Primer in Black Holes, Mach's Principle and
Gravitational Energy}, Nova Science, New York.

Berman,M.S. (2008a) - \textit{A General Relativistic Rotating Evolutionary
Universe, }Astrophysics and Space Science, \textbf{314, }319-321.

Berman,M.S. (2008b) - \textit{A General Relativistic Rotating Evolutionary
Universe - Part II, }Astrophysics and Space Science, \textbf{315}, 367-369.
Posted with another title, in Los Alamos Archives
http://arxiv.org/abs/0801.1954 .

Gomide, F.M.(1963) - Nuovo Cimento, \textbf{30}, 672.

Leighton, R.B. (1959) - \textit{Principles of Modern Physics,} McGraw-Hill, N.Y.

\end{document}